\documentclass{DISproc}
\usepackage{bm, cite}
\begin{document}

\title{Forward Drell-Yan plus backward jet as a test of  BFKL evolution}

\author{{\slshape Martin Hentschinski$^1$, Clara Salas$^2$}\\[1ex]
$^1$Physics Department, Brookhaven National Laboratory, Upton, 11973, USA \\ \\ 
$^2$Instituto de F{\' i}sica Te{\' o}rica  
UAM/CSIC,   C/ Nicol\'as Cabrera 13-15,   \\ Universidad Aut{\' o}noma de Madrid, Cantoblanco, E-28049 Madrid, Spain
}

\contribID{xy}

\doi  

\maketitle

\begin{abstract}
  We study Drell-Yan plus jet events where the gauge boson is produced
  in the forward direction of one of the colliding protons and a jet
  is produced in the forward direction of the second proton. The
  resulting large rapidity difference between the final states then
  opens up the phase space for BFKL evolution.  First numerical results on partonic level are provided. 
\end{abstract}

\section{Introduction}

Due to its large center of mass energy the LHC allows for the study of
forward physics using methods of perturbative QCD.  Among them we find
forward production of different systems such as high $p_T$ jets
\cite{Deak:2011ga}, heavy quark pairs ~\cite{Salas:prep} and Drell-Yan
(DY) processes where a virtual photon or $Z$ boson decays into a pair
of leptons~\cite{Marzani:2008uh, Hautmann:2012sh}. The study of these
type of processes is interesting as they allow to probe parton
distribution functions at very small values of $x$ which have not been
reached in so-far collider experiments. They therefore provide a
possibility to test formalisms which have been especially developed
for the description of small $x$ processes and which go beyond the
standard formulation in terms of collinear factorization by including
additional small $x$ enhanced contributions. The starting point of
such studies is given by BFKL evolution which resums small $x$
logarithms on the level of partonic scattering amplitudes at leading
logarithmic (LL)~\cite{Fadin:1975cb} and next-to-leading-logarithmic
(NLL)~\cite{Fadin:1998py} accuracy. In addition, since small $x$
evolution ultimatively leads to high parton densities, such processes
may further allow for the observation of saturation effects which
require an extension of the BFKL formalism, see {\it e.g.}
\cite{Betemps:2001he}.

While studies of inclusive observables provide strong hints for the
presence of BFKL evolution in small $x$ data, see {\it e.g.}
\cite{Hentschinski:2012kr}, a proper identification of relevant
effects at small $x$ is  hard to achieve. Cancellations between
different final states minimize the sensitivity to the particular
feature of the employed method and deviations from inclusive evolution
equations due to small $x$ effects may be partly hidden into the
chosen initial conditions. It is therefore necessary to turn to the
study of more exclusive observables   to distinguish different
effects at small $x$.  Among
these exclusive observables there is a class of events where the
entire dependence of the process on the non-perturbative dynamics is
treated within conventional collinear factorization. These observables
typically involve hard events in the forward region of both scattering
protons, while the large difference in rapidity between the hard final
states opens up the phase space for BFKL evolution. Among the best
explored processes of this type are `Mueller-Navelet' jets which
consist of a high $p_T$-jets in the forward regions of each
proton. Currently this process is one of the few examples where a
complete NLL BFKL description exists~\cite{Colferai:2010wu}. In contrast to na\"ive expectations, the result
reveals a strong dependence on the next-to-leading order corrections to
the jet impact factors. At the same time, the numerical differences
between the NLL resummed result and a pure collinear NLO result remain rather
small for a large class of observables.

This observation motivates the study of a new type of forward-backward
observable, where a DY pair is produced in the forward direction of
one of the particles instead of a jet.  The hope is that this
observable is able to better distinguish between standard NLO results
and NLL BFKL resummed predictions and allows to identify universal features of BFKL evolution.  Even though the large virtuality
of the photon and/or the mass of the $Z$ diminishes at first the value
of the strong coupling constant $\alpha_s$, the rapidity difference
between lepton pair and backward jet remains large at the Large Hadron
Collider, $\Delta Y < 7$ and a study of BFKL evolution appears
meaningful.  In addition, study of new final states may also trigger
new theoretical efforts for an improved definition of impact factors
and lead to the identification of new BFKL observables. In the
following we present some partial results of our study, which are
currently restricted to the partonic level. For  details we refer to our paper in preparation~\cite{HS_prep}.

\section{The leading-order DY impact factor}
\label{sec:LODYimpact}

\begin{figure}[h]
  \centering
  \parbox{4cm}{\includegraphics[width = 4cm]{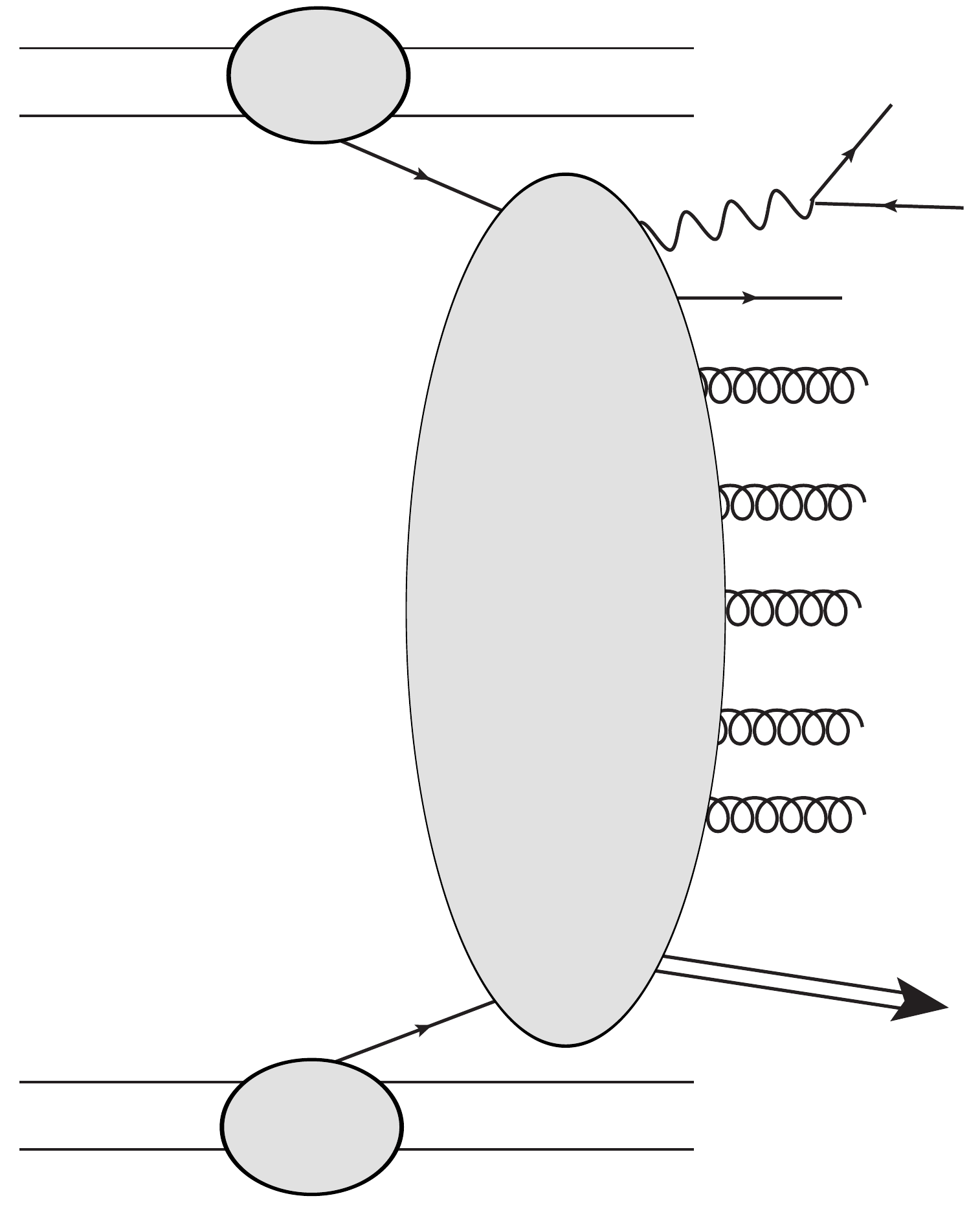}} \parbox{.3cm}{$\,$}
\parbox{3cm}{\includegraphics[height = 1.4cm]{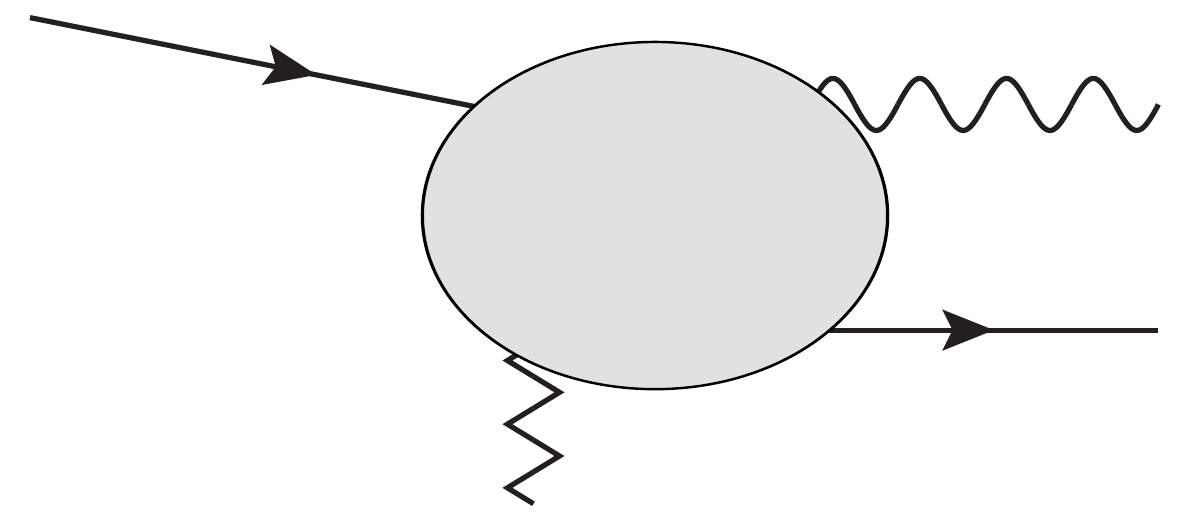}} $\, =\, $
\parbox{2cm}{\includegraphics[height = 1.4cm]{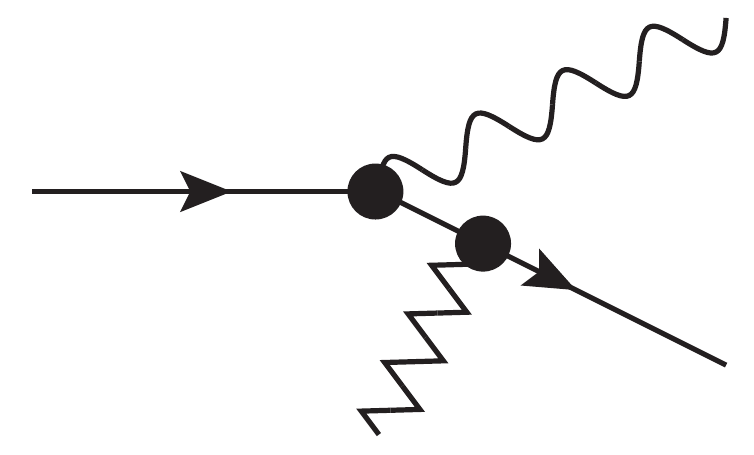}} $\, + \,$
\parbox{2cm}{\includegraphics[height = 1.4cm]{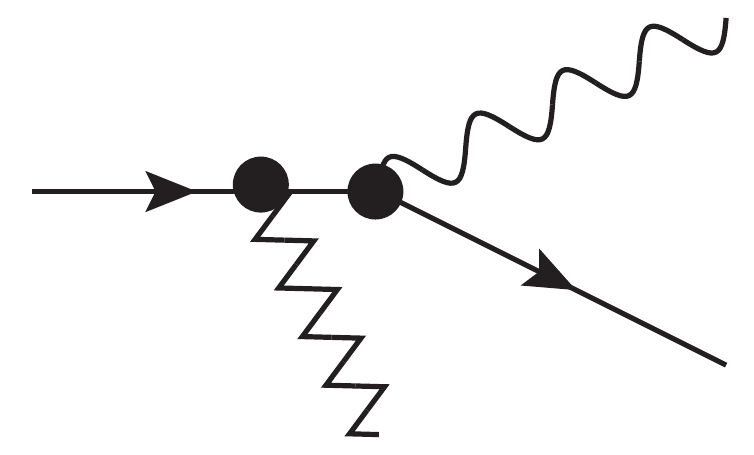}}

\parbox{4cm}{\center (a)} \parbox{7.5cm}{\center (b)}
  \caption{a) A large difference in rapidity between the forward gauge boson ($\gamma^*, Z$) and the backward jet opens up the phase space for BFKL evolution. b) The leading order DY impact factor is obtained as the sum of two effective diagrams where the $t$-channel gluon carries eikonal polarizations.}
  \label{fig:DYXsec}
\end{figure}
In the current study we restrict to the LO impact factor, where
relevant diagrams can be found in Fig.~\ref{fig:DYXsec}.b. A complete
NLO study seems possible using Lipatov's effective
action~\cite{Lipatov:1995pn} which is currently explored at
NLO~\cite{Hentschinski:2011tz}. The leading order impact factor reads
\begin{align}
  \Phi_{Zq}&   
 = 
 \frac{c_f\alpha_s \sqrt{N_c^2 - 1}}{   \pi {\bm k}^2  N_c}  
 \left [ \frac{z {\bm k}^2 \big((1-z)^2 + 1 \big)   + 2 M^2 (1-z)z  } {D_1 D_2}  - \frac{M^2 z(1-z)}{D_1^2}  - \frac{M^2 (1-z)z}{D_2^2}  \right] \notag \\
& \quad D_1  =  ({\bm q} - z {\bm k})^2 + (1-z)M^2 \qquad \qquad  D_2 =  {\bm q}^2  +  (1-z)M^2
\end{align}
Here $M$ denotes the mass of the $Z$ boson and the virtuality of the
photon respectively, while $c_f$ yields the coupling of the gauge
boson to the quark. ${\bm q}$ and ${\bm k}$ are the transverse momenta
of the final state gauge boson and initial gluon, while $z$ is the
momentum fraction of the initial quark momentum carried on by the
gauge boson.

\section{Preliminary numerical results at partonic level}
\label{sec:num}

The above impact factor carries a logarithmic singularity if the final
state quark turns to be soft.  After convolution of the impact factor with the BFKL Green's function, this corresponds to the limit $z \to 1$.  To avoid this singularity we study
ratios of angular coefficients $\mathcal{C}_n = \langle \cos n \phi
\rangle$, where $\phi$ denotes the azimuthal angle between the jet and
the gauge boson.
\begin{figure}[h!]
  \centering
\parbox{6.5cm}{\center \includegraphics[height = 4cm]{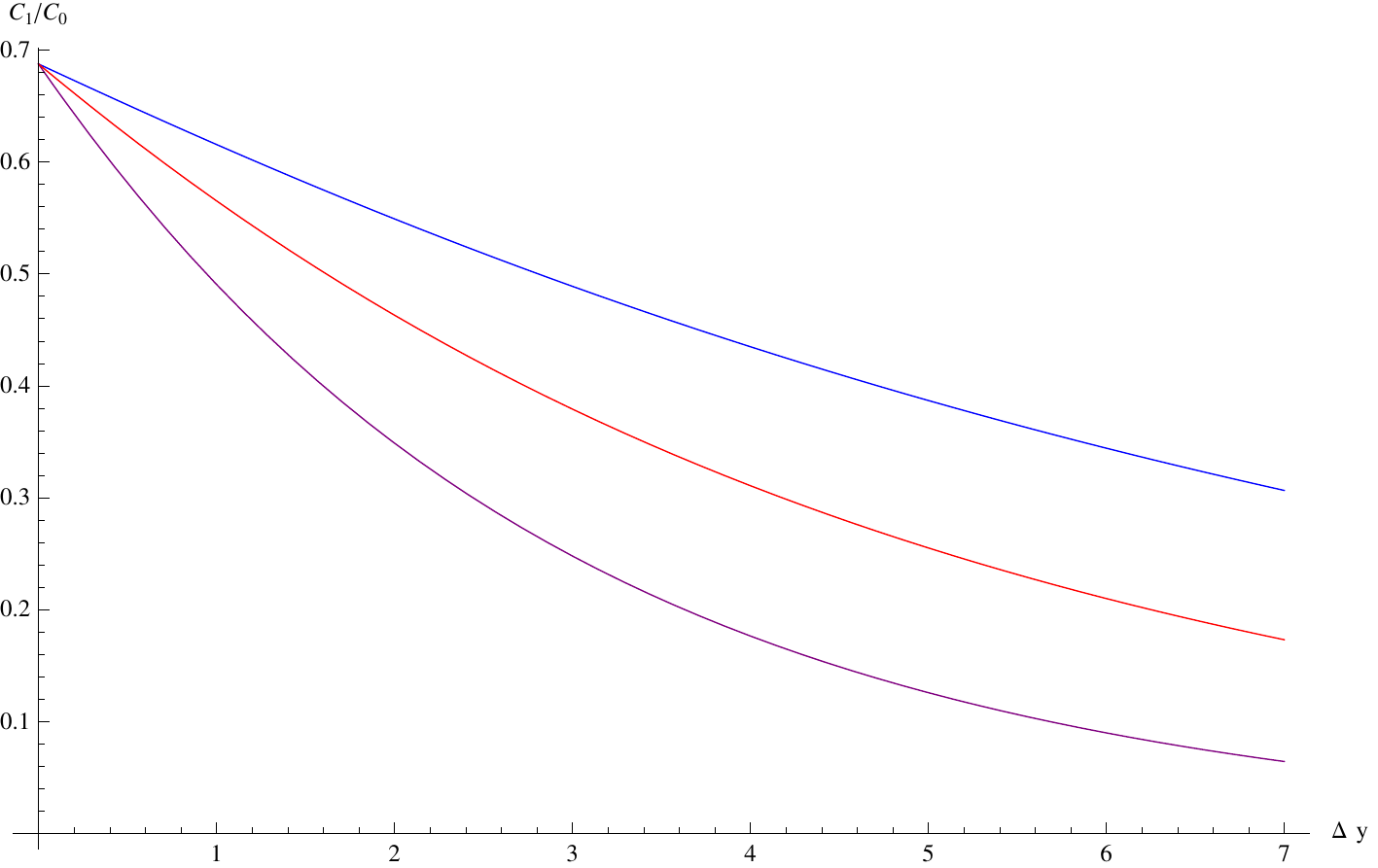}}
\parbox{6.5cm}{\center \includegraphics[height = 4cm]{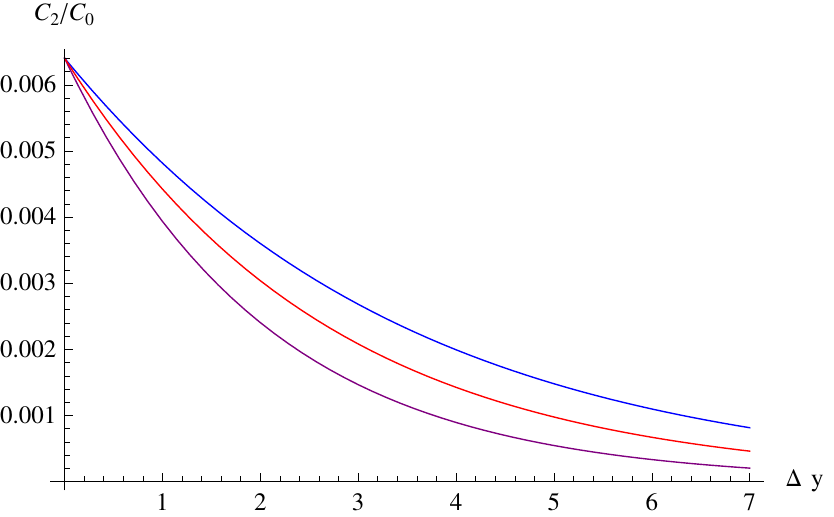}}
\parbox{6.5cm}{\center (a) \\ }\parbox{6.5cm}{\center (b) \\ }

\parbox{6.5cm}{\center \includegraphics[width = 6.5cm]{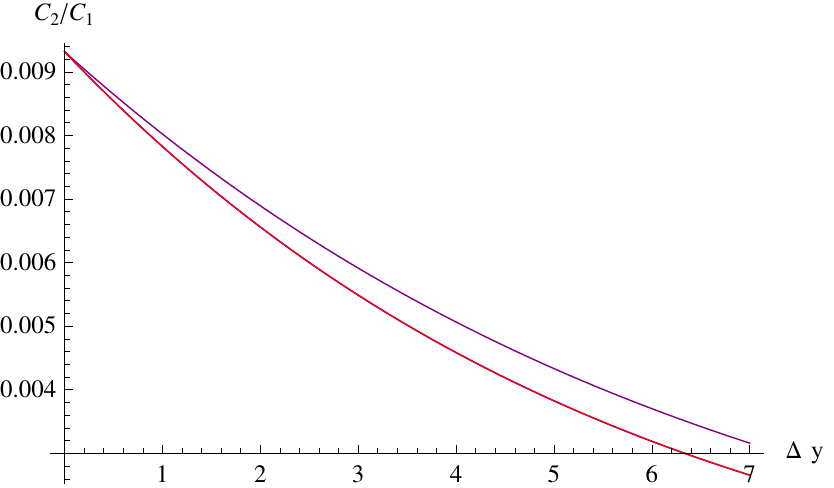}}

\parbox{6.5cm}{\center (c)}

\caption{Ratios of angular coefficients $\mathcal{C}_1/\mathcal{C}_0$, $\mathcal{C}_2/\mathcal{C}_0$
  and $\mathcal{C}_2/\mathcal{C}_1$ versus $\Delta y =| y_Z -
  y_{\text{jet}}|$ on partonic level with $\alpha_s(M_Z^2)$ and $z =
  0.9$. The transverse momentum of the $Z$-boson is taken to be ${\bm
    q} = 2$ GeV while jets with transverse momenta ${\bm p > 20}$
  GeV are considered.  The dependence on $\Delta y$ is described by
  the LL (purple), NLL (blue) and NLL RG improved (see~\cite{Vera:2005jt} for details)
  (red) BFKL Green's function }
  \label{fig:numericalb}
\end{figure}


In a preliminary study, see Fig.~\ref{fig:numericalb}, which restricts to the partonic level and
conformal part of the NLO BFKL kernel, we use a fixed value $z =0.9$
while the QCD coupling is taken at the $Z$-scale. As in the jet-jet
case we find that the best convergence of the BFKL prediction is
achieved for ratios which do not contain the angular coefficient
$\mathcal{C}_0$. A complete study at hadronic level faces additional
complications. Even though formally finite, the limit $z \to 1$, which is
naturally encountered once  the convolution with parton distribution
functions is included, leads to large contributions which can even cause 
negative results for some of the angular coefficients.  The study of
these effects and their appropiate treatment as well as the inclusion of the running coupling corrections of the NLO BFKL Green's function  is currently in progress.

\section*{Acknowledgments}
We are grateful for financial support from  the MICINN under grant FPA2010-17747, the
Research Executive Agency (REA) of the European Union under the Grant
Agreement number PITN-GA-2010-264564 (LHCPhenoNet) and Comunidad de Madrid (HEPHACOS ESP-1473). M.H. further  acknowledges support from the German Academic
Exchange Service (DAAD) and   the U.S. Department of Energy under contract number DE-AC02-98CH10886 and a BNL ``Laboratory Directed Research and Development'' grant (LDRD 12-034).
 

{\raggedright
\begin{footnotesize}



\end{footnotesize}
}


\end{document}